\documentclass[epj]{svjour}
\usepackage{graphics}
\usepackage{amsmath}
\newcommand{\ket}[1]{\vert #1 \rangle}
\newcommand{\bra}[1]{\langle #1 \vert}

\newcommand{\braket}[2]{\langle #1 \vert #2 \rangle}

\newcommand{\ketbra}[2]{\vert #1 \rangle \langle #2 \vert}
\begin{document}
\title{Exact results on decoherence and entanglement in a system of N driven atoms and a dissipative cavity mode}
\author{M. Bina \and F. Casagrande \thanks{e-mail: federico.casagrande@mi.infn.it}\and A. Lulli
}                     
\institute{Dipartimento di Fisica, Universit\`{a} di Milano, via
Celoria 16, 20133 Milano, Italy\\}
\date{Received: date / Revised version: date}
%
\abstract{We solve the dynamics of an open quantum system where N strongly driven
two-level atoms are equally coupled on resonance to a dissipative
cavity mode. Analytical results are derived on decoherence,
entanglement, purity, atomic correlations and cavity field mean
photon number. We predict decoherence-free subspaces for the
whole system and the N-qubit subsystem, the monitoring of quantum
coherence and purity decay by atomic populations measurements,
the conditional generation of atomic multi-partite entangled
states and of cavity cat-like states. We show that the dynamics of atoms prepared in states invariant under
permutation of any two components remains restricted within the subspace
spanned by the completely symmetric Dicke states. We discuss examples and applications in the cases
$N=3,4$.
} 
\authorrunning{M. Bina \emph{et al.}}
\titlerunning{Decoherence and entanglement in a system of N driven atoms and a dissipative cavity mode}
\maketitle
\section{Introduction}\label{intro}
Cavity quantum electrodynamics (QED) concerns the interaction of
atoms (or ions) with a quantized radiation field in a microwave or
optical
cavity~\cite{Haroche_Raimond,ReviewOptical,ReviewOptical2}. The
basic physical principles are quite well understood \cite{JC} and
accurately tested~\cite{JCexp,JCexp2,JCexp3}. Impressive advances
in the experimental control on these systems allow to study
fundamental issues in quantum physics such as entanglement and
decoherence, related to the nonlocal correlations of composite
systems at the microscopic level and to the boundary between the
quantum and the classical descriptions. These issues have raised a
huge interest due to the potentialities of the peculiar quantum
behavior for applications in quantum information processing,
communication, and computation~\cite{NielsenChuang} and the
necessity to protect quantum coherence from noisy
environments~\cite{decoh,decoh2}. In this framework it is
important to investigate nontrivial solvable models, representing
somewhat idealized versions of systems implementable with the
present cavity QED technology. In this paper we exactly solve the
dynamics of an open multipartite system, where $N$ two-level atoms
are equally coupled on resonance to a dissipative cavity mode and coherently
driven by a strong external field. Cooled, trapped,
deterministically loaded atoms
~\cite{trapped-atoms,trapped-atoms2,trapped-atoms3} and trapped
ion systems~\cite{ioncavity,ioncavity2} in optical cavities, as
well as Rydberg atoms crossing microwave cavities \cite{Haroche_Raimond},
appear as the most promising candidates for implementations.\\
We present a quite compact solution of the open system dynamics
derived by phase-space techniques~\cite{Glauber2}, used in previous works~\cite{SDOAL,Bina}.
The N-atom subsystem can be described as a pseudo-spin
system where the independent coupling of each atom to the cavity
combines with the invariance of dynamics under permutation of any
two atoms. A peculiarity of the system is that this description
does not hold in the standard (energy or computational) basis, but
in a rotated one. The permutational invariance reflects in the
atomic coupling to the environment, leading to the existence of
both global and atomic decoherence-free subspaces
(DFS)~\cite{DFS,DFS2}. In the latter case an initial $N$-qubit
entanglement remains protected and available e.g. for quantum
memories or quantum processors~\cite{Moehring,Zoller}. The
structure of the general solution allows predicting a way to
monitor the decay of quantum coherence and purity by
measurements of atomic probabilities. In the limit of unitary
dynamics these measurements can conditionally generate mesoscopic
cat-like states of the cavity field. The preparation of atoms in
states which are invariant under the exchange of any atom pair
restricts the atomic dynamics to the subspace spanned by the
completely symmetric Dicke states~\cite{Mandel-Wolf}, some of
which are genuinely multipartite entangled states~\cite{Solano}.
This further simplifies the description of system dynamics in the important cases
of atoms all prepared in a same state.
Selected results for three and four atoms provide further insight
on system and subsystems
dynamics, including  preservation and conditional generation of multipartite entanglement.\\
In Sect. \ref{sec:2} we introduce the model. The analytical
solution of system dynamics is derived Sect. \ref{sec:3}, where
applications to $N=3,4$ are also reported. Results for transient
and steady state regimes are discussed in Sect. \ref{sec:4}, and
multipartite entanglement protection is the object of Sect. \ref{sec:5}. The
main results are summarized in the Conclusions.

\section{The physical model}\label{sec:2}
We consider a set of $N$ two-level atoms interacting with a
dissipative cavity field mode. The transition frequency $\omega_a$
between excited and ground states, $\ket{e}_l$ and $\ket{g}_l$
($l=1,...,N$), is the same for all the $N$ atoms. A coherent
external field of frequency $\omega_a$ simultaneously drives the
atoms during the interaction with the cavity mode of frequency
$\omega_f$~\cite{Solano-Agarwal,Lougovski}.  This kind of system
is feasible in cavity QED experiments with two-level Rydberg atoms
in a microwave cavity~\cite{Haroche_Raimond} or with three-level
atoms effectively reduced to two levels interacting with an
optical cavity~\cite{SDOAL}, due to relevant advances recently
achieved in cooling, trapping and deterministically loading atoms
in optical cavities~\cite{trapped-atoms}. In both regimes atomic
decays can be neglected, as we shall assume from now on. Similar
dynamics could be also implemented by trapped ions interacting
with a
cavity mode~\cite{ioncavity}.\\
The whole system Hamiltonian is
\begin{equation}\begin{split}\label{eq:hamiltonianSDJC}
\hat{\mathcal{H}}(t)&=\hbar\omega_f
\hat{a}^{\dag}\hat{a}+\hbar\sum_{l=1}^N\Big [
\frac{\omega_a}{2}\hat{\sigma}_{z,l} +
g(\hat{\sigma}_l^{\dag}\hat{a}+\hat{\sigma}_l\hat{a}^{\dag}) +\\
&+\Omega( e^{-i\omega_a t}\hat{\sigma}_l^{\dag}+e^{i\omega_a
t}\hat{\sigma}_l)\Big ] ,
\end{split}\end{equation}
where $\Omega$ is the Rabi frequency associated
with the coherent driving field amplitude, $g$ the atom-cavity
mode coupling constant taken equal for all atoms, $\hat{a}$
($\hat{a}^{\dag}$) the field annihilation (creation) operator,
$\hat{\sigma}_l=\ket{g}_l\bra{e}$
($\hat{\sigma}_l^{\dag}=\ket{e}_l\bra{g}$) the atomic lowering
(raising) operator, and
$\hat{\sigma}_{z,l}=\ket{e}_l\bra{e}-\ket{g}_l\bra{g}$ the
inversion operator.\\
In the perspective of experimental implementation of our scheme we
add the effects of cavity mode dissipation, while we focus on
resonance conditions ($\omega_a=\omega_f$) in order to derive an
analytical solution of the system dynamics.\\
Therefore, we must solve the following master equation (ME) for
the statistical density operator $\hat{\rho}_N'$ of the whole
system
\begin{equation}\label{eq:ME1}
\dot{\hat{\rho}}_N'=-\frac{i}{\hbar}[\hat{\mathcal{H}},\hat{\rho}_N']
+ \hat{\mathcal{L}}_f\hat{\rho}_N',
\end{equation}
where
\begin{equation}
\hat{\mathcal{L}}_f\hat{\rho}_N'=\frac{k}{2}[2\hat{a}\hat{\rho}_N'
\hat{a}^{\dag}-\hat{a}^{\dag}\hat{a}\hat{\rho}_N' - \hat{\rho}_N'
\hat{a}^{\dag}\hat{a}]
\end{equation}
is the standard Liouville superoperator which describes the
dissipative decay of the cavity field mode, with the rate $k$, due
to the coupling to a thermal bath at zero temperature. \\
In the interaction picture the dissipative terms remain unchanged
and the ME (\ref{eq:ME1}) can be rewritten as
\begin{equation}\label{eq:MEIP}
\dot{\hat{\rho}}_N^I=-\frac{i}{\hbar}[\hat{\mathcal{H}}^I,\hat{\rho}_N^I]
+ \hat{\mathcal{L}}_f\hat{\rho}_N^I
\end{equation}
where the Hamiltonian (\ref{eq:hamiltonianSDJC}) has been replaced
by the time-independent Hamiltonian
$\hat{\mathcal{H}}^I=\hat{\mathcal{H}}_0+\hat{\mathcal{H}}_1$ with
\begin{align}\label{eq:IP_H}
\hat{\mathcal{H}}_0&= \hbar\Omega\sum_{l=1}^N\left (
\hat{\sigma}_l^{\dag}+ \hat{\sigma}_l\right ),
&\hat{\mathcal{H}}_{1}=\hbar g\sum_{l=1}^N\left (
\hat{\sigma}_l^{\dag}\hat{a}+\hat{\sigma}_l\hat{a}^{\dag}\right ).
\end{align}
Now we consider the unitary transformation
$\hat{\mathcal{U}}(t)=e^{\frac{i}{\hbar}\hat{\mathcal{H}}_0 t }$
and we derive for the density operator
$\hat{\rho}_N=\hat{\mathcal{U}}\hat{\rho}_N^I\hat{\mathcal{U}}^{\dag}$
the following ME:
\begin{equation}\label{eq:ME2}
\dot{\hat{\rho}}_N=-\frac{i}{\hbar}[\hat{\mathcal{H}}_1',\hat{\rho}_N]
+ \hat{\mathcal{L}}_f\hat{\rho}_N
\end{equation}
where the transformed Hamiltonian
$\hat{\mathcal{H}}_1'=\hat{\mathcal{U}}\hat{\mathcal{H}}_1\hat{\mathcal{U}}^{\dag}$
can be written as
\begin{equation}\begin{split}
\hat{\mathcal{H}}_1'(t)&=\frac{\hbar g}{2}\hat{a}\sum_{l=1}^N\Big
[ (1-e^{-2i\Omega t})\hat{\sigma}_l +(1+e^{2i\Omega
t})\hat{\sigma}_l^{\dag}\Big ] + {\rm h.c.}
\end{split}\end{equation}
In the strong-driving regime for the interaction between the atoms
and the external coherent field, $\Omega\gg g$, we can use the
rotating-wave approximation obtaining the effective Hamiltonian
\cite{Solano-Agarwal,Lougovski}
\begin{equation}\label{eq:effective_H}
\hat{\mathcal{H}}_{\rm eff}= \frac{\hbar g}{2}(\hat{a}
+\hat{a}^{\dag})\sum_{l=1}^N(\hat{\sigma}^{\dag}_l+\hat{\sigma}_l)
\end{equation}
We notice the presence of Jaynes-Cummings
($\hat{\sigma}_j^{\dag}\hat{a} +\hat{\sigma}_j\hat{a}^{\dag}$) as
well as anti-Jaynes-Cummings ($\hat{\sigma}_j^{\dag}\hat{a}^{\dag}
+\hat{\sigma}_j\hat{a}$) coupling terms of each coherently driven
atom with the cavity field. Hereinafter we shall solve the master
equation for the whole system density operator $\hat{\rho}_N(t)$
\begin{equation}\label{eq:ME}
\dot{\hat{\rho}}_N=-\frac{i}{\hbar}[\hat{\mathcal{H}}_{\rm
eff},\hat{\rho}_N] + \hat{\mathcal{L}}_f\hat{\rho}_N.
\end{equation}

\section{Analytical solution and system dynamics} \label{sec:3}
In order to solve the general $N$-atom problem we introduce the
collective atomic operator
$\hat{S}_x=\frac{1}{2}\sum_{l=1}^N\hat{\sigma}_{x,l}=\frac{1}{2}\sum_{l=1}^N(\hat{\sigma}^{\dag}_l+\hat{\sigma}_l)$,
so that the effective Hamiltonan assumes the simple form
\begin{equation}
\hat{\mathcal{H}}_{\rm eff}=\hbar g (\hat{a} +\hat{a}^{\dag})
\hat{S}_x.
\end{equation}
We recall that the eigenstates of the spin operator
$\hat{\sigma}_{x,l}$ are the rotated states
$\ket{\pm}_l=\frac{\ket{g}_l\pm\ket{e}_l}{\sqrt{2}}$ where
$\hat{\sigma}_{x,l}\ket{\pm}_l=\lambda_l^{\pm}\ket{\pm}_l$ with
$\lambda_l^{\pm}=\pm 1$. For the whole atomic subspace we consider
the basis of $2^N$ states $\{\ket{i}_N\}$ where any
$\ket{i}_N$ is an eigenstate of the collective spin operator
$\hat{S}_x$. The corresponding eigenvalue
$s_i=(1/2)\sum_{l=1}^N\lambda_l^{\pm}$ is half the difference
between the number of $\ket{+}$ and $\ket{-}$ components of state
$\ket{i}_N$, regardless of the exchange of any qubit pair, and it
can assume $N+1$ values from $-N/2$ to $N/2$ with steps $|\Delta
s_i|=1$. We notice that the eigenvalues $s_i$ have a degeneracy
order given by
$n(s_i)=\frac{N!}{(N/2+s_i)!(N/2-s_i)!}$ that is greater than one if $-N/2<s_i<N/2$. \\
The general solution of the ME (\ref{eq:ME}) can be derived by
introducing the decomposition of the density operator
$\hat{\rho}_N(t)=\sum_{i,j=1}^{2^N} {}_N\bra{i}\hat{\rho}_N(t)\ket{j}_N\ket{i}_N\bra{j}$
on the N-atom rotated basis $\ket{i}_N$, so that
it is equivalent to the following set of $2^{2N}$ uncoupled
evolution equations for the field operators
$\hat{\rho}_{N,ij}={}_N\bra{i}\hat{\rho}_N(t)\ket{j}_N$
\begin{equation}\begin{split}\label{eq:ME_elements}
\dot{\hat{\rho}}_{N,ij}&=
-ig \big [ s_i(\hat{a}+\hat{a}^{\dag})\hat{\rho}_{N,ij}-s_j\hat{\rho}_{N,ij}(\hat{a}+\hat{a}^{\dag}) \big ]\\
&+\hat{\mathcal{L}}_f\hat{\rho}_{N,ij}.
\end{split}\end{equation}
Equation~(\ref{eq:ME_elements}) can be solved by a combination of
phase space techniques~\cite{Glauber2} with the method of
characteristics ~\cite{Barnett}. Starting from the cavity in the
vacuum state $\ket{0}$ and the atoms in any pure state
\begin{equation}\label{eq:initial_state} \ket{\Psi(0)}_N=
\ket{0}\otimes\sum_{i=1}^{2^N}c_{N,i}\ket{i}_N
\end{equation}
with the normalization condition $\sum_{i=1}^{2^N}|c_{N,i}|^2=1$,
we obtain the compact solution for the whole system density
operator
\begin{equation}\begin{split}\label{eq:rhoN}
\hat{\rho}_N(t)=&\sum_{i,j=1}^{2^N}c_{N,i}c_{N,j}^*[f(t)]
^{(s_i-s_j)^2} \times\\
&\times\ketbra{-2s_i\alpha(t)}{-2s_j\alpha(t)}
\otimes\ket{i}_N\bra{j}.
\end{split}\end{equation}
We see that the dynamics correlates the eigenstates of $\hat{S}_x$ with
cavity field coherent states of amplitude proportional to
\begin{equation}\label{eq:alpha(t)}
\alpha(t)=i\frac{g}{k}\left ( 1-e^{-\frac{k}{2}t}\right ).
\end{equation}
The one-atom decoherence function
\begin{equation}
f(t) = f_1(t)e^{2|\alpha(t)|^2}=e^{
-\frac{2g^2}{k}t+\frac{4g^2}{k^2}\left ( 1-e^{-\frac{k}{2}t}\right
) }e^{2|\alpha(t)|^2} \label{eq:f(t)}
\end{equation}
naturally splits into two parts: $f_1(t)$,
which will appear in the atomic subsystem dynamics, and
$e^{2|\alpha(t)|^2}$, that is the field states normalization. It
is responsible for the decay of coherences and depends on the
dimensionless parameters $(g/k) ^2$ and
$kt$.\\
In order to evaluate the degree of mixedness of the state
$\hat{\rho}_N(t)$ we derive from eq.~(\ref{eq:rhoN}) the purity
\begin{equation}\begin{split}\label{eq:purity_tot}
\mu_N(t)&=Tr[\hat{\rho}_N^2(t)]\\
&=\sum_{i,j=1}^{2^N}|c_{N,i}|^2|c_{N,j}|^2[f(t)]^{2(s_i-s_j)^2}.
\end{split}\end{equation}
A remarkable consequence of the general solution of
eq.~(\ref{eq:rhoN}) is the existence of a global DFS for any even
value of the number N of atoms, when the eigenvalue $s_i$ can
assume the value zero. In this case there is no time
evolution for the initial states of eq.~(\ref{eq:initial_state})
containing only the corresponding $n(0)=N!/[(N/2)!]^2$ atomic
eigenstates $\ket{i}_N$. Let us consider for example the case of
$N=4$ atoms. The DFS is spanned by the tensor product of the
cavity vacuum state and $n(0)=6$ states $\ket{i}_4$ with the same
number of $\ket{+}\text{ and }\ket{-}$ components (see Table
\ref{tab:1}). It preserves any initial global state within this
subspace, protecting any entangled atomic
preparation for quantum information purposes.\\
Another interesting feature of the case with even N follows from
the presence in eq.~(\ref{eq:rhoN}) of terms with the cavity field
in the vacuum state. Namely, if the optical cavity field is
accessible to measurements, the absence of a response by an on/off
detector generates a pure N-qubit state, that can be a
multipartite entangled state. We consider again the $N=4$ atoms
case. Starting e.g. from the four atoms prepared in the ground
state, the density operator of eq.~(\ref{eq:rhoN}) contains a
time-independent part $(\sqrt{6}/4)(\ket{0}\otimes\ket{\Psi}_a)$,
where
$\ket{\Psi}_a=(1/\sqrt{6})(\ket{++--}+\ket{+-+-}+\ket{+--+}+\ket{-++-}+\ket{-+-+}+\ket{--++})$.
Hence a null measurement of the optical cavity field generates the
pure 4-qubit state $\ket{\Psi}_a$ whose entanglement properties
will be discussed later.\\
\begin{table*}
\caption{Atomic DFSs for $N=4$ qubits.}
\label{tab:1}       
\begin{center}
\begin{tabular}{ccl}
\hline\noalign{\smallskip}
$\mathbf{s_i}$ & $\mathbf{n(s_i)}$ & $\mathbf{\ket{i}_4}$ \\
\noalign{\smallskip}\hline\hline\noalign{\smallskip}
2 & 1 & $\{ \ket{++++}\}$ \\
1 & 4 & $\{ \ket{+++-}$,$\ket{++-+}$,$\ket{+-++}$,$\ket{-+++}\}$ \\
0 & 6 & $\{ \ket{++--}$,$\ket{+-+-}$,$\ket{+--+}$,$\ket{-+-+}$,$\ket{-++-}$,$\ket{--++}\}$ \\
-1 & 4 & $\{ \ket{---+}$,$\ket{--+-}$,$\ket{-+--}$,$\ket{+---}\}$  \\
-2 & 1 & $\{ \ket{----}\}$ \\
\noalign{\smallskip}\hline
\end{tabular}
\end{center}
\end{table*}

\subsection{Completely symmetric Dicke states}
\begin{figure}
\centering
\resizebox{0.4\textwidth}{!}{\includegraphics{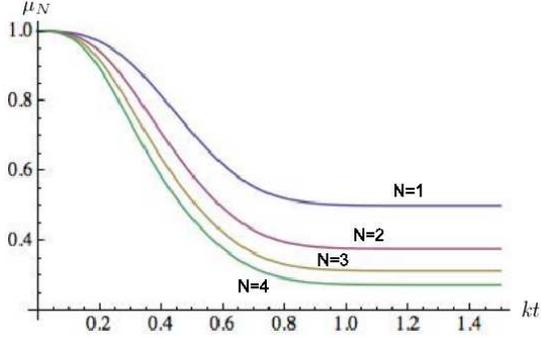}}
\caption{Time evolution of the purity of the whole system for different values of $N$ and
for the fixed dimensionless parameter $g/k=5$.}
\label{fig:purityN}       
\end{figure}

In order to exploit the system dynamical invariance under exchange
of any atom pair, we consider initial states
(eq.~(\ref{eq:initial_state})) having in the atomic part only
symmetric states or symmetrized combinations of states
$\ket{i}_N$. In this case the atomic part of $\hat{\rho}_{N}(t)$
remains confined in the subspace spanned by only $N+1$ (instead of $2^N$) states that we
denote as $\ket{\frac{N}{2},s}$, where $-\frac{N}{2}\leq
s\leq\frac{N}{2}$ with steps $|\Delta s| = 1$. We notice that the above states are
analogous to the so-called
completely symmetric Dicke states (CSD) in \cite{Mandel-Wolf}, written in the energy basis
instead of the rotated one. For instance for $N=4$ the states $\ket{2,s}$ with $-2\leq
s\leq 2$ are the symmetrized combinations of the states listed in Table 1. All the previous
treatment can be adapted correspondingly. In particular, starting
from any superposition of CSD states
\begin{equation}\label{eq:initial_state_CSD} \ket{\Psi(0)}_N=
\ket{0}\otimes\sum_{s=-N/2}^{N/2}b_{N,s}\ket{\frac{N}{2},s}
\end{equation}
with the normalization condition
$\sum_{s=-N/2}^{N/2}|b_{N,s}|^2=1$, the general
solution~(\ref{eq:rhoN}) can be rewritten as
\begin{equation}\begin{split}\label{eq:rhoN_CSD}
\hat{\rho}_N(t)=&\sum_{s,s'=-N/2}^{N/2}b_{N,s}b_{N,s'}^*[f(t)] ^{(s-s')^2}\times\\
&\times\ketbra{-2s\alpha(t)}{-2s'\alpha(t)}
\otimes\ketbra{\frac{N}{2},s}{\frac{N}{2},s'}.
\end{split}\end{equation}
In this case the interaction correlates cavity field coherent
states with atomic CSD states. These results include the important
case of all atoms prepared in the ground state, where
$b_{N,s}=(1/\sqrt{2^N})n(s)$. In another relevant case, with all
atoms prepared in the excited state, the only change
in~(\ref{eq:rhoN_CSD}) is the replacement $f(t)\rightarrow -f(t)$.
In such cases the purity (\ref{eq:purity_tot}) reduces to
\begin{equation}\label{eq:purity(t)_gg}
\mu_N(t)=\frac{1}{2^{2N}}\sum_{s,s'=-N/2}^{N/2}f(t)^{2(s-s')^2}
\end{equation}
whose asymptotic value can be written in a closed form in terms of
the gamma function $\Gamma$
\begin{equation}\label{eq:puritySS}
\mu_N^{SS}=\frac{1}{2^{2N}}\sum_{s=-N/2}^{N/2}n^2(s)=\frac{1}{2^{2N}}\sum_{l=0}^{N}\binom{N}{l}^2=
\frac{\Gamma(N+\frac{1}{2})}{\sqrt{\pi}\Gamma(N+1)}
\end{equation}
where $l\equiv s+N/2$ and $\binom{N}{l}$ is the binomial coefficient.
In Fig. \ref{fig:purityN} we show the time evolution of the system purity (\ref{eq:purity(t)_gg})
for a fixed value of $g/k$ and different qubit numbers $N=1,...,4$, where the steady state values
are $\mu_1^{SS}=1/2$, $\mu_2^{SS}=3/8$, $\mu_3^{SS}=5/16$, $\mu_4^{SS}=35/128$. The greater
the value of $N$, the faster the decay of the global coherences. Varying the ratio $g/k$
instead of $N$, the asymptotic behavior does not change whereas the decay is faster (slower)
for increasing (decreasing) values of $g/k$.\\
We remark that the atomic preparation in one of the CSD states is
equivalent to the qubit encoding in the corresponding DFS. In particular,
for even values of $N$, the CSD atomic state with $s=0$  and the cavity
in the vacuum state belong to a global DFS. \\

\subsection{Subsystem dynamics}
\begin{figure*}
\begin{center}
\begin{minipage}[c]{.40\textwidth}
\centering
\resizebox{0.7\textwidth}{!}{\includegraphics{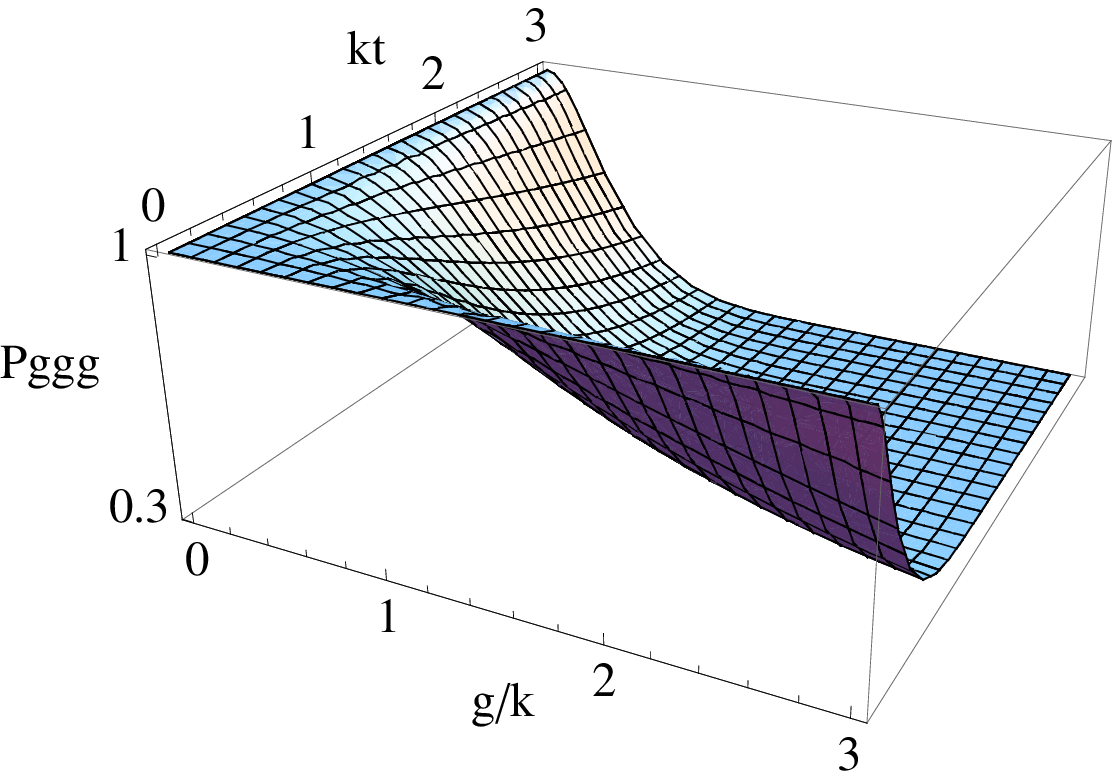}}
\end{minipage}%
\hspace{10mm}%
\begin{minipage}[c]{.40\textwidth}
\centering
\resizebox{0.7\textwidth}{!}{\includegraphics{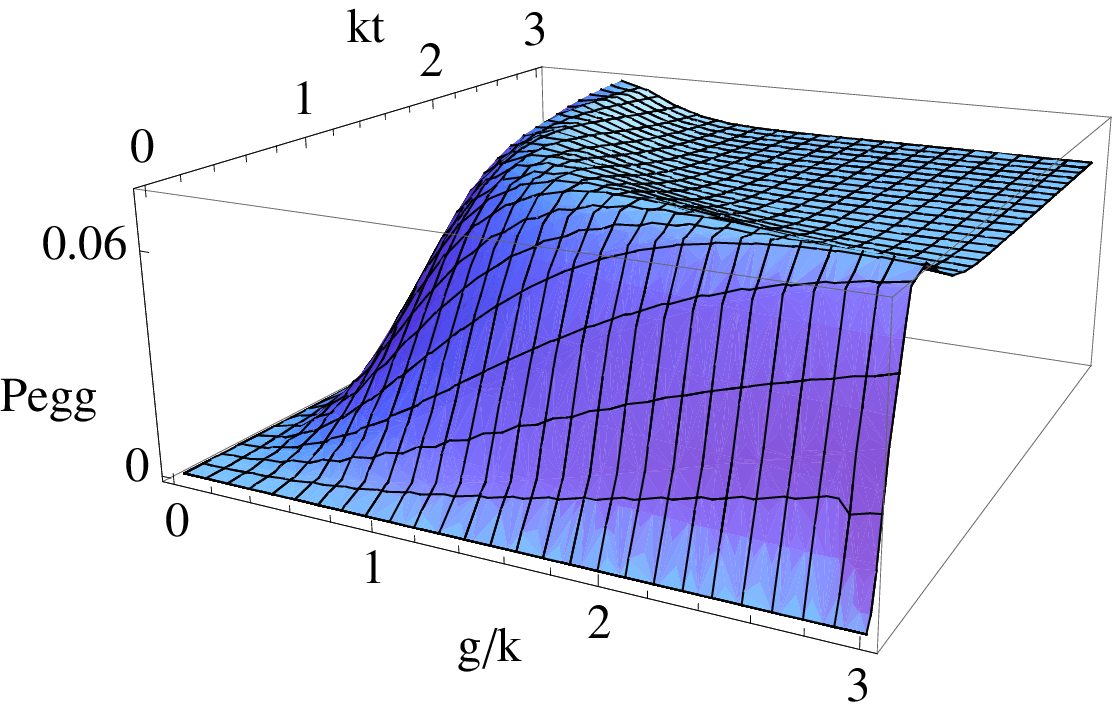}}
\end{minipage}
\hspace{10mm}%
\begin{minipage}[c]{.40\textwidth}
\centering
\resizebox{0.7\textwidth}{!}{\includegraphics{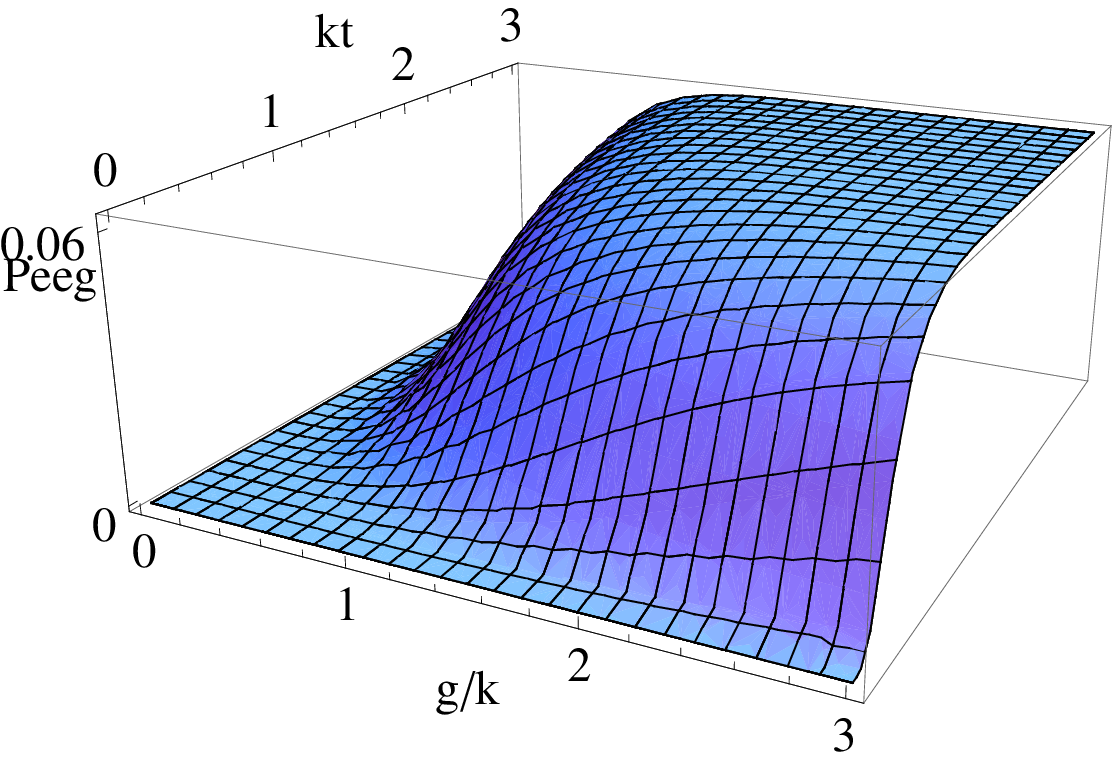}}
\end{minipage}\hspace{10mm}%
\begin{minipage}[c]{.40\textwidth}
\centering
\resizebox{0.7\textwidth}{!}{\includegraphics{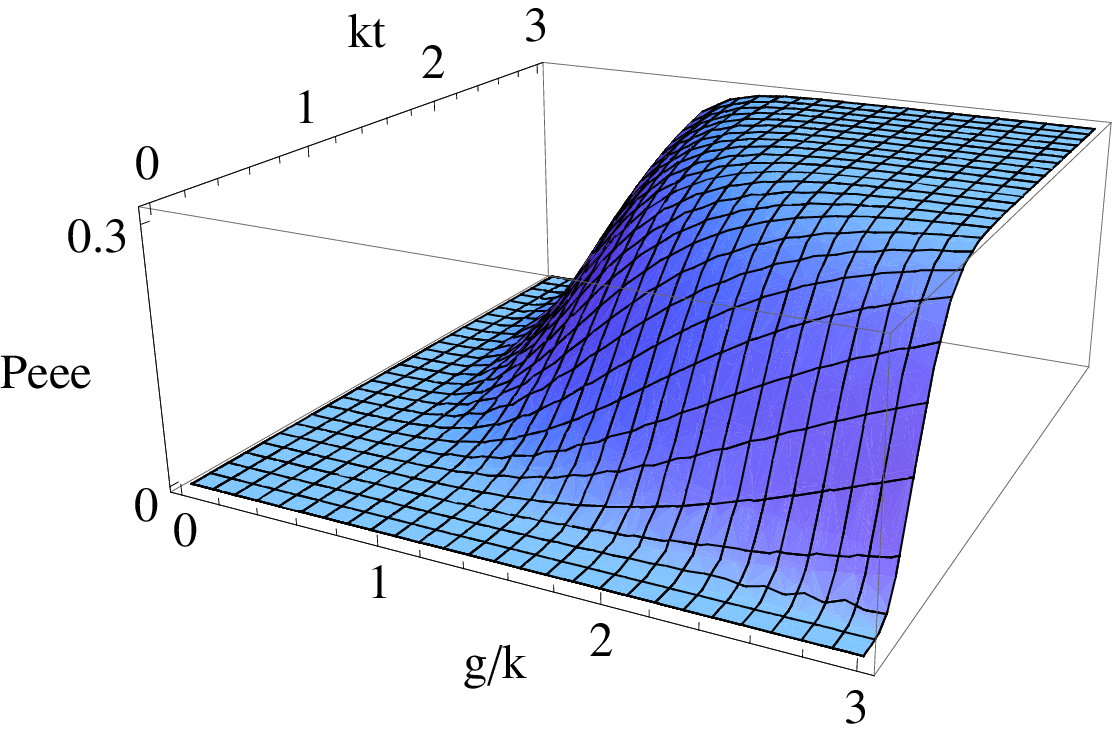}}
\end{minipage}
\caption{Three-atom joint probabilities vs dimensionless coupling
constant $g/k$ and time $kt$ from
eqs.~(\ref{eq:joint_prob}).\label{fig:prob}}
\end{center}
\end{figure*}

We derive some general results on the cavity mode and the atomic
subsystems, providing the time-dependent expressions of the
corresponding reduced density operators and purities, as well as
of the mean number of photons in the
cavity.\\
If we trace the whole system density operator of
eq.~(\ref{eq:rhoN}) over the atomic variables, we find the
expression for the reduced density operator of the cavity field
\begin{equation}\label{eq:rho_f}
\hat{\rho}_{N,f}(t)=\sum_{i=1}^{2^N}|c_{N,i}|^2
\ketbra{-2s_i\alpha(t)}{-2s_i\alpha(t)}
\end{equation}
that is a statistical mixture of coherent states, and
whose purity is
\begin{equation}\begin{split}\label{eq:purity_f}
\mu_{N,f}(t)&=Tr[\hat{\rho}_{N,f}^2(t)]=\\
&=\sum_{i,j=1}^{2^N}|c_{N,i}|^2|c_{N,j}|^2[e^{2|\alpha(t)|^2}]^{2(s_i-s_j)^2}.
\end{split}\end{equation}
From the density operator $\hat{\rho}_{N,f}(t)$ we
can derive the expression of the mean photon number
\begin{equation}\label{eq:photon_number}
\langle \hat{a}^{\dag}\hat{a} \rangle (t)
=Tr_f[\hat{\rho}_{N,f}(t)\hat{a}^{\dag}\hat{a}]
=4|\alpha(t)|^2\sum_{i=1}^{2^N}s_i^2|c_{N,i}|^2.
\end{equation}
In the case of all atoms in the ground state,
$c_{N,i}=1/\sqrt{2^N}$, one obtains
\begin{equation}
\langle \hat{a}^{\dag}\hat{a}\rangle (t)=\frac{|\alpha(t)|^2}{2^{N-2}}\sum_{l=0}^{N}\left( l-\frac{N}{2} \right ) ^2\binom{N}{l}=N|\alpha(t)|^2
\end{equation}
showing that each atom gives the same
average contribution to the
cavity field.\\
By tracing the whole system density operator over the field
variables, we obtain the reduced atomic density operator
\begin{equation}\label{eq:rho_a}
\hat{\rho}_{N,a}(t)=\sum_{i,j=1}^{2^N}c_{N,i}c_{N,j}^*[f_1(t)]^{(s_i-s_j)^2}\ket{i}_N\bra{j}.
\end{equation}
We notice that if the atoms are prepared in any superposition of
eigenstates $\ket{i}_N$ corresponding to a degenerate eigenvalue
$-N/2<s_i<N/2$, the state does not evolve. Therefore we identify
$N-1$ atomic DFSs with dimension $n(s_i)$ greater than one, where an initial entanglement can be protected.
The purity of state (\ref{eq:rho_a}) is
\begin{equation}\begin{split}\label{eq:purity_a}
\mu_{N,a}(t)&=Tr[\hat{\rho}_{N,a}^2(t)]=\\
&=\sum_{i,j=1}^{2^N}|c_{N,i}|^2|c_{N,j}|^2[f_1(t)]^{2(s_i-s_j)^2}.
\end{split}\end{equation}
The decay of atomic purity ruled by $f_1(t)$ is faster than the global purity decay
(\ref{eq:purity_tot}), ruled by $f(t)$. However the asymptotic behavior is the same
and we can use the result (\ref{eq:puritySS}) in order to make a comparison with
the case of maximally mixed states, whose purity is equal to $1/2^N$.
\begin{figure}
\centering
\resizebox{0.4\textwidth}{!}{\includegraphics{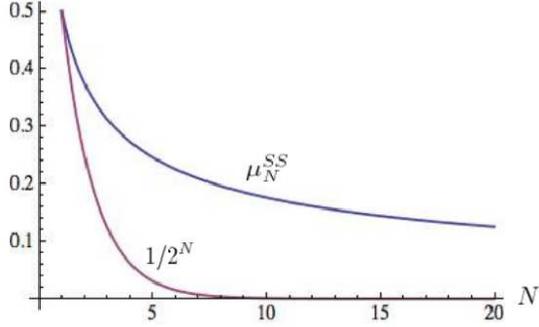}}
\caption{Comparison between the $N$-qubit purity $\mu_{N,a}(t)$ in (\ref{eq:purity_a}) evaluated at steady state
and $1/2^N$ that is the purity of a maximally mixed state.
}
\label{fig:confronto}       
\end{figure}
In Fig. \ref{fig:confronto}
we see that the state is maximally mixed only for $N=1$, where actually
the atom becomes maximally entangled with cavity field \cite{SDOAL}. For
any $N>1$ the state is never maximally mixed due to the survival of coherences
in the DFSs. Also we notice that the field purity (\ref{eq:purity_f}) remains slightly
larger than the atomic one because the decoherence function $f_1(t)$
is replaced by a non-vanishing exponential function.\\
As an application of the atomic subsystem dynamics we rewrite the
atomic density matrix eq.~(\ref{eq:rho_a}) in the standard basis
for the case $N=3$. Starting, for instance, from the three atoms
in the ground state the diagonal matrix elements provide the
following joint probabilities for the atomic level populations
\begin{subequations}\label{eq:joint_prob}\begin{align}
P_{eee}(t)&=\frac{1}{32}\Big [ 10-15f_1(t)+6f_1^4(t)-f_1^9(t)\Big
]\label{eq:p_eee_3}\\
P_{eeg}(t)&=\frac{1}{32}\Big  [ 2-f_1(t)-2f_1^4(t)+f_1^9(t)\Big  ]\label{eq:p_eeg_3}\\
P_{egg}(t)&=\frac{1}{32}\Big  [ 2+f_1(t)-2f_1^4(t)-f_1^9(t)\Big  ]\label{eq:p_egg_3}\\
P_{ggg}(t)&=\frac{1}{32}\Big  [ 10+15f_1(t)+6f_1^4(t)+f_1^9(t)\Big
]\label{eq:p_ggg_3}
\end{align}\end{subequations}
where eqs.~(\ref{eq:p_eeg_3}) and (\ref{eq:p_egg_3}) represent one
third of the probability to detect, respectively, two atoms in the
excited state or in the ground state, independently from the
atomic ordering. The three-atom probabilities
(\ref{eq:joint_prob}) are shown in Fig. \ref{fig:prob}.

We can see that at steady state the joint probability that three
atoms are in the same state is equal to $5/16$, that is an atomic
correlation (bunching) effect, whereas the joint probability of
the other two outcomes is $3/16$, showing an antibunching
effect. By exploiting the above expressions (\ref{eq:joint_prob})
it is possible to monitor the two-atom decoherence function
$f_1^4(t)$ measuring the sums $P_{eee}(t)+P_{ggg}(t)$ or
$P_{eeg}(t)+P_{egg}(t)$. Remarkably the
N-qubit decoherence originates from the one-atom
decoherence function ~\cite{SDOAL} which can be
monitored by atomic population measurements via the relation
$f_1(t)=P_g(t)-P_e(t)$, as well as the N-qubit purity according to (\ref{eq:purity_a}).
In that case the atom and the field can
approach maximally entangled states in the limits $kt\ll1$ and
$(g/k)^2\gg 1$, and the entanglement (measured by the Von Neumann
subsystem entropy) is also described by
$f_1(t)$.

\section{Transient and steady state results} \label{sec:4}
\begin{figure}[h!]
\centering
\resizebox{0.4\textwidth}{!}{\includegraphics{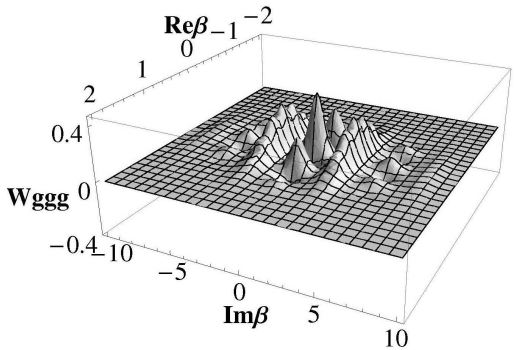}}
\caption{Wigner function $W_{ggg}$ of the cavity field state conditioned to
the detection of the three atoms in the ground state, for
parameters values $kt=0.05\text{ and }g/k=110$.}
\label{fig:2}       
\end{figure}

In the Hamiltonian limit, $kt\ll 1$, of small cavity decay rate
and/or short interaction times, $f(t)\rightarrow 1$, $\alpha(t)\rightarrow
\tilde{\alpha}(t)\equiv i\frac{gt}{2}$ and $\rho_N(t)\rightarrow
\ket{\tilde{\Psi}(t)}_N \bra{\tilde{\Psi}(t)}$ where the global cat-like
state
\begin{equation}\label{eq:psiN}
\ket{\tilde{\Psi}(t)}_N=\sum_{i=1}^{2^N}c_{N,i}
\ket{-2s_i\tilde{\alpha}(t)}\otimes\ket{i}_N.
\end{equation}
As an example, for $N=3$ we consider the generation of the pure
state $\ket{\tilde{\Psi}(t)}_3$. Starting from
three atoms in the ground state, so that $c_{3,i}=1/\sqrt{8}$, we
obtain for $kt\ll 1$ an evolved state that we rewrite in the standard atomic basis
\begin{equation}\begin{split}\label{eq:psi3}
&\ket{\tilde{\Psi}}_3=\frac{1}{\sqrt{8}}\Big [(\ket{-3\tilde{\alpha}}-3\ket{-\tilde{\alpha}}+3\ket{\tilde{\alpha}}-\ket{3\tilde{\alpha}})\otimes\ket{eee}\\
&+(\ket{-3\tilde{\alpha}}+3\ket{-\tilde{\alpha}}+3\ket{\tilde{\alpha}}+\ket{3\tilde{\alpha}})\otimes\ket{ggg}\\
&+(\ket{-3\tilde{\alpha}}-\ket{-\tilde{\alpha}}-\ket{\tilde{\alpha}}+\ket{3\tilde{\alpha}})\otimes(\ket{eeg}+\ket{ege}+\ket{gee})\\
&+(\ket{-3\tilde{\alpha}}+\ket{-\tilde{\alpha}}-\ket{\tilde{\alpha}}-\ket{3\tilde{\alpha}})\otimes(\ket{egg}+\ket{geg}+\ket{gge})
\Big ]
\end{split}\end{equation}
where for brevity we have defined
$\tilde{\alpha}\equiv\tilde{\alpha}(t)$.
We notice a superposition of mesoscopic cat-like states of the
cavity field correlated with atomic states with the same number of
ground (or excited) atoms, which are two fully separable and two
entangled 3-qubit states (a W and an inverted-W state)~\cite{Dur}.
An interesting consequence of eq.~(\ref{eq:psi3}) is that a
simultaneous detection of the three atoms in any state prepares
the cavity field in the corresponding cat-like state. In Fig. \ref{fig:2} we show
the Wigner function that describes in phase space the cat-like state
generated for atomic detections in the ground state. \\
After the transient the coupling of the field to the environment
introduces in the solution (\ref{eq:rhoN}) the field-atoms
coherences $f(t), f^4(t),...,f^{N^2}(t)$. Note that these powers
of the decoherence function can be obtained by the substitution
$g\rightarrow Ng$, which exactly reflects the independent
interaction of each atom with the cavity field.\\
In the steady state limit $kt\gg 1$ the density operator
$\hat{\rho}_N(t)$ becomes a statistical mixture of the pure states
superimposed in the global cat-like state (\ref{eq:psiN})
generated in the transient
\begin{equation}\label{eq:rhoN_SS}
\hat{\rho}_N^{SS}=\sum_{i=1}^{2^N}|c_{N,i}|^2
\ketbra{-2s_i\alpha^{SS}}{-2s_i\alpha^{SS}}
\otimes\ket{i}_N\bra{i},
\end{equation}
where $\alpha^{SS}=i\frac{g}{k}$. The system (subsystem) purity
at steady state was discussed in the previous section.\\

\section{Protection of multipartite entanglement}\label{sec:5}
Now we recall some concepts and tools
in order to analyze the multipartite entanglement properties of
some atomic states encoded in DFSs. The 3-tangle measure introduced in \cite{Coffman}
evaluates the amount of entanglement shared by all the three qubits through the quantity
$\tau_{123}=C_{12}^2+C_{13}^2 - C_{1(23)}^2$, where $C_{ij}$ is
the concurrence of the qubit pair $(i,j)$. A generalization to the case
of N qubits (with N even) was given in \cite{Wong} by the
$N$-tangle measure defined as $\tau_N=|\braket{\psi}{\tilde{\psi}}|^2$
with $\ket{\tilde{\psi}}=\sigma_y^{\otimes_N}\ket{\psi^*}$, where $\ket{\psi}$ is the generic
$N$-qubit state, $\ket{\psi^*}$ its complex conjugate and $\sigma_y$ one of the Pauli matrices. Another useful
tool is the residual bipartite entanglement measure (see \cite{Dur}) which evaluates
the robustness of entanglement against the loss of information; this
measure is provided, for instance, by the average squared concurrence
$\overline{C^2}$ calculated for any two residual qubits
when the other $N-2$ are traced out.\\
Following the previous analysis concerning the DFSs we first
consider the atomic subsystem for $N=3$. In this case we find that
the three qubits do not evolve in time if they are prepared in any
of the four decoherence-free CSD states
\begin{subequations}\begin{align}
&\ket{3/2,3/2} =\ket{+++}\label{eq:CSD_3/2}\\
&\ket{3/2,1/2} =\frac{\ket{++-} + \ket{+-+} + \ket{-++}}{\sqrt{3}}\label{eq:CSD_1/2}\\
&\ket{3/2,-1/2}=\frac{\ket{+--} + \ket{-+-} + \ket{--+}}{\sqrt{3}}\label{eq:CSD_-1/2}\\
&\ket{3/2,-3/2}=\ket{---}\label{eq:CSD_-3/2}.
\end{align}\end{subequations}
Two of them, (\ref{eq:CSD_3/2}) and (\ref{eq:CSD_-3/2}), are
manifestly separable. The other two states, (\ref{eq:CSD_1/2}) and (\ref{eq:CSD_-1/2}), show
interesting entanglement properties. They have no full tripartite
entanglement ($\tau_{123}=0$) according to the $3$-tangle
measure. However each qubit pair retains the maximal residual bipartite
entanglement $\overline{C^2}=4/9$. These kind
of states show a multipartite entanglement characteristic of
W-like states.\\
Let us now investigate the dynamics of four qubits which
presents five decoherence-free CSD states, including two separable
states $\ket{2,\pm 2}$, and the multipartite entangled states $\ket{2,\pm 1}$ and
$\ket{2,0}$. The relevance for applications in quantum information
processing is that the state $\ket{2,0}$ turns out to be maximally
entangled according to the $4$-tangle measure ($\tau_4=1$), whereas
the states $\ket{2,\pm 1}$ have no four-partite entanglement ($\tau_4=0$), but
each of them exhibits an equal maximal reduced bipartite entanglement, $\overline{C^2}=1/4$ (W-like states), by
tracing over any qubit pair. We remark that all CSD states of the type
$\ket{N/2,\tilde{s}}$ with $\tilde{s}=\pm (N-2)/2$ have entanglement properties
similar to that of states $\ket{W_N}$ introduced in \cite{Dur}.
By tracing the atomic density operators $\ketbra{N/2,\tilde{s}}{N/2,\tilde{s}}$ over any $N-2$ parties we
always obtain the reduced density operators for the bipartite system
\begin{equation}
\rho_{\pm}=\frac{1}{N}\left ( 2\ketbra{\Phi^-}{\Phi^-}+(N-2)\ketbra{\pm\pm}{\pm\pm}\right ).
\end{equation}
Hence for the average squared concurrence we simply obtain the value $\overline{C^2}=(2/N)^2$.\\
We further notice that we can rewrite the states $\ket{2,\pm 1}$ and $\ket{2,\pm 0}$
as
\begin{equation}\begin{split}\label{eq:phi_simm}
\ket{2,\pm
1}&=\frac{1}{\sqrt{2}}(\ket{\pm\pm}_{12}\ket{\Phi^-}_{34}+\ket{\Phi^-}_{12}\ket{\pm\pm}_{34})\\
\ket{2,0}&=\frac{1}{\sqrt{6}}\Big (
\ket{\Phi^-}_{12}\ket{\Phi^-}_{34}+\ket{\Phi^-}_{13}\ket{\Phi^-}_{24}+\\
&+\ket{\Phi^-}_{14}\ket{\Phi^-}_{23}\Big )
\end{split}\end{equation}
thus generalizing the
results derived in~\cite{Bina} where we showed that the two-atom maximally entangled Bell state
$\ket{\Phi^-}_{ij}=\frac{1}{\sqrt{2}}(\ket{+-}_{ij}+\ket{-+}_{ij})$ and
the two separable states $\ket{\pm\pm}$ do not evolve in time.\\
Another interesting application is to encode the four qubits in
some states of a special basis called Bell gem \cite{Jaeger}, which
is a generalization of the well known Bell basis
$\ket{\Phi^\pm}=(1/\sqrt{2})(\ket{gg}\pm \ket{ee})$ and
$\ket{\Psi^\pm}=(1/\sqrt{2})(\ket{ge}\pm \ket{eg})$. It is
composed by maximally entangled states, according to the
$4$-tangle measure ($\tau_4=1$), which can be obtained by simple quantum logic
circuits starting from four unentangled qubits in the
computational basis. Let us consider the cavity field prepared in
the vacuum state $\ket{0}$ and the four qubits in one of the last
three elements of the Bell gem
$(1/\sqrt{2})(\ket{\Phi^+\Psi^+}-\ket{\Psi^+\Phi^+})$,
$(1/\sqrt{2})(\ket{\Psi^-\Phi^-}\pm\ket{\Phi^-\Psi^-})$. The whole
system does not evolve in time because these three initial atomic
states belong to the DFS corresponding to the eigenvalue $s_i=0$,
thus maintaining the maximum multipartite entanglement in the
atomic subsystem.

\section{Conclusions}\label{sec:6}
We have solved the dynamics of a feasible cavity QED system where
$N$ strongly driven two-level atoms are equally and resonantly
coupled to an optical field mode in contact with an environment
and initially in the vacuum state. For negligible atomic decay we
have derived a compact solution of the open system master equation
in terms of coherent field states, atomic pseudo-spin states, and
suitable decoherence functions. We have derived and discussed a
number of exact results on system and subsystems dynamics which
are also of interest for quantum information applications,
including decoherence-free subspaces and multipartite ($N$-qubit)
entanglement protection. In addition we have suggested a way to
monitor decoherence by atomic population measurements. For atoms
prepared in symmetric states with respect to the exchange of any
atom pair, including the physically important preparation in the
same (ground or excited) state, the dynamics is entirely expressed
in terms of symmetric Dicke states. Applications in the cases with
$N=3,4$ have been discussed.


%

\end{document}